\documentclass[conference]{IEEEtran}
\IEEEoverridecommandlockouts
\usepackage{amsmath,amsfonts}
\usepackage{algorithmic}
\usepackage{algorithm}
\usepackage{array}
\usepackage[caption=false,font=scriptsize]{subfig}
\usepackage{textcomp}
\usepackage{stfloats}
\usepackage{url}
\usepackage{verbatim}
\usepackage{graphicx}
\usepackage{cite}
\usepackage{hhline}
\usepackage{bm}
\usepackage{color}
\usepackage{longtable}	
\usepackage{multicol}
\usepackage{float}
\usepackage{multirow}
\usepackage{xspace}
\usepackage{cite}
\usepackage{amsmath,amssymb,amsfonts}
\usepackage{algorithmic}
\usepackage{tikz}
\usetikzlibrary{shapes.geometric, arrows}
\usepackage{textcomp}
\usepackage{diagbox}
\usepackage{xcolor}
\usepackage{graphicx}
\usepackage{nicefrac}
\usepackage{textcmds}
\usepackage{bm}
\usepackage{amsmath,amssymb}
\usepackage{enumerate}
\usepackage{graphicx}
\usepackage{tikz}
\usetikzlibrary{shapes.geometric, arrows}
\usepackage{hhline}
\usepackage{longtable}	
\usepackage{nicefrac}
\usepackage{multicol}
\usepackage{float}
\usepackage{multirow}
\usepackage{epsf,psfrag,amssymb,amsfonts,color,cite,fancybox}
\usepackage{array}
\usepackage[mathscr]{eucal}
\usepackage{algorithmic}
\tikzstyle{startstop} = [rectangle, rounded corners, minimum width=1.2cm, minimum height=0.8cm,text centered, draw=black, fill=white!20]
\tikzstyle{io} = [trapezium, trapezium left angle=70, trapezium right angle=110, minimum width=0cm, minimum height=0cm, text centered, draw=black, fill=red!30]
\tikzstyle{process} = [rectangle, minimum width=0cm, minimum height=0cm, text centered, draw=black, fill=orange!15, rounded corners]
\tikzstyle{decision} = [diamond, minimum width=1.2cm, minimum height=0.8cm, text centered, draw=black, fill=blue!10]
\tikzstyle{arrow} = [thick,->,>=stealth]

\usepackage{textcomp}
\usepackage{diagbox}
\usepackage{xcolor}
\usepackage{soul}
\usepackage{enumitem}
\allowdisplaybreaks

\usepackage{optidef}
\usepackage{mathtools}
\newtagform{rred}{\color{rred}(}{)}
\begin{document}

\title{An Optimal Energy Management Algorithm Considering Regenerative Braking and Renewable Energy for EV Charging in Railway Stations\\}

\author{Georgia Pierrou,~\IEEEmembership{Member,~IEEE,} Yannick Zwirner, and Gabriela Hug,~\IEEEmembership{Senior Member,~IEEE}
\thanks{This work is supported by ETH Mobility Initiative under MI-GRANT 2020-HS-396. }
\thanks{Georgia Pierrou and Gabriela Hug are with the Power Systems Laboratory, Department of Information Technology and Electrical Engineering, ETH Zurich, Zurich, 8092, Switzerland. (email: gpierrou@ethz.ch, ghug@ethz.ch). \color{black}Yannick Zwirner is with SBB Energy, 3000 Bern, Switzerland (email: yannick.zwirner@sbb.ch).
}

}

\maketitle

\begin{abstract}
This paper proposes a novel optimal Energy Management System (EMS) algorithm for Electric Vehicle (EV) charging in smart electric railway stations with renewable generation. As opposed to previous railway EMS methods, the proposed EMS coordinates the combined Regenerative Braking Energy (RBE), renewable generation, electric railway demand and EV charging demand at the EV parking lot of the railway station. Numerical results using a scenario-based approach on an actual railway station in Chur, Switzerland demonstrate that the proposed
algorithm can effectively minimize the expected daily operating cost for the train station over an entire year. 
\end{abstract}


\section{Introduction}
% Dynamic load modeling plays an important role in power system stability. Traditionally, load dynamics have been considered as the driving parameters towards voltage instability \cite{Cutsem}. Inaccurate load models may lead to misconceptions regarding the stability limits and control decisions \cite{Makarov96}. To ensure reliability in power system operation and to avoid costly system controller design, it is essential that the load dynamic behavior is accurately captured and load characteristics are properly identified. However, load identification remains a challenging issue due to the large number, the complexity and the time variability of the different load components as well as the uncertainty brought about by poor measurements and customer behavior \cite{Bai09}.

Transportation electrification, including the electrification of private and commercial vehicles, is expected to have an increasing impact on power system operation. The central location of electric railway stations
with electric buses and parking lots offering the option for \qq{park and rail} provides an excellent opportunity to transform existing railway infrastructure into major energy hubs. Indeed, such vision, taking also advantage of renewable generation, Regenerative Braking Energy (RBE) capabilities and energy storage systems (ESS) to improve the efficiency and enable energy savings, has been discussed at North
American and European railway operator entities \cite{NA,EU,Ratni14}.  Therefore, it is important to conduct systematic studies and develop
appropriate Energy Management Systems (EMS) to coordinate the aforementioned concepts and optimize system operation under this new framework.

To address the energy analysis and management of electric railway systems with any ESS, RBE and renewable generation different approaches have been proposed in the literature. 
In \cite{7482762}, a multi-period optimal power flow problem formulation is presented to analyze the effect of renewable uncertainty and ESS for planning purposes and improve railway system efficiency and energy savings. A Mixed Integer Linear Programming (MILP) model of a railway EMS concept covering ESS, RBE, PV, and different
pricing schemes is described in \cite{Sengor18}. A methodology to optimize the sizing of ESS in electric railway systems while considering \color{black}power produced from Regenerative Braking (RB) \color{black} is proposed in \cite{6877707}. In \cite{7028816}, a fuzzy logic supervision strategy to integrate
renewable generation and ESS in a railway
substation is investigated. However, EV charging was not considered in these aforementioned works.

%  The weighted least squares method has been applied in \cite{Lin93} to derive the load model. In \cite{Guo12}, the nonlinear least square (NLS) technique is used to identify the parameters of a dynamic recovery load in the presence of voltage changes due to load tap changer operation. An improved NLS method is proposed in \cite{Zhang17}, however, an initial guess %priori information 
% on parameter values %\color{red}{What is the priori information?}\color{black} 
% is required for the estimation. A hybrid learning technique that combines the Genetic Algorithm and the Levenberg-Marquard algorithm for load parameter estimation has been proposed in \cite{Bai09}. However, the implementation of the aforementioned optimization-based methods leads to high computational burden, making the real-time application prohibitive \cite{Rouhani16}. 

Limited work has been done on integrating EV charging in railway operation. The authors in \cite{5589545} study the design of a sustainable urban mobility system including a metro transit system, EVs and hybrid vehicles. Nevertheless, renewable generation is not considered in this analysis. A framework to utilize RBE, ESS and renewable generation to meet the demand of the EV parking lot at a railway station is implemented in \cite{9745034}. However, the variation of renewables is not included in the scenarios, which could influence the charging decisions. Also, the concept of selling the excess power back to the main grid is not considered in the cost function. In \cite{7587629}, a method that demonstrates the viability of providing
renewable power charging services for EVs is presented by analyzing the impact of the ESS capacity. Yet, the proposed algorithm considers renewable generation only for charging purposes, whereas railway demand is not taken into account.

In this paper, we propose a novel EMS algorithm that integrates RBE, ESS, renewable generation and EV charging for the optimal operation of electric railway stations. The main contributions of the paper are as follows:
\vspace{-3pt}
\begin{enumerate}
    \item  The proposed approach incorporates the variations of local renewable generation availability in the set of scenarios.
    %\item The method applies in the presence stochastic load fluctuations. A main advantage of the algorithm is that no knowledge on the statistical properties (e.g variance) of the uncertain parameters is required, in contrast to \cite{Wangxz:2017}.
    \item The potential of selling any excess power back to the main grid is exploited as part of the objective function.
    \item To the best of the authors' knowledge, this work represents the first attempt to implement an optimal energy management strategy in order to meet the combined electric railway demand and EV charging demand. %exploiting the properties of recursion, an effective and computationally inexpensive online algorithm can be developed and integrated for real time estimation purposes.  
\end{enumerate}
\vspace{-3pt}
The remainder of the paper is organized as follows: Section
II introduces the mathematical formulation for the scenario-based energy management model. In Section III, the proposed EMS algorithm including the estimation of the EV charging demand is elaborated. Section IV provides a
numerical study that demonstrates the effectiveness
of the method. Section V presents the conclusions and perspectives for future work.

\section{Energy Management Model}

\subsection{Objective Function}
In this paper, the minimization of the expected daily operating cost for the railway station consumption across all scenarios is considered as the main objective. Therefore, the objective function of the EMS model is selected as follows:
\begin{equation} 
 \underset {} 
 {\text{min}}  \sum_s \sum _t \pi_s \hspace{0.025in} (C_{G}^{t,s}P_G^{t,s}-C_{S}^{t,s}P_S^{t,s})\Delta t 
 \label{eq:objfunction}
\end{equation}
where $\pi_s$ denotes the probability of each scenario; $P_G^{t,s}$ is the power bought from the main grid to meet the demand; $P_S^{t,s}$ is the power sold back to the main grid; $C_{G}^{t,s}$ is the time dependent purchasing electricity price; $C_{S}^{t,s}$ is the time dependent selling electricity price; $\Delta t$ is the time step. Costs related to investment decisions are not within the scope of this work.

\subsection{PV Generation Modeling}
In addition to the conventional generation from the main grid, it is assumed that photovoltaic (PV) generation is installed at the main train station level. To model the PV generated power, solar radiation measurements that are available locally may be used. Briefly speaking, a piece-wise function determines the active power corresponding to the available solar radiation as follows \cite{Pierrouu19}:
\begin{equation}
\label{eq:solar_p_val}
P_{PV}^{t,s}={P}_{PV}^{t,s}(\beta^{t,s})=\left\{ \begin{array}{*{35}{l}}
\displaystyle \frac{{{\beta^{t,s}}^{2}}}{{{r}_{c}}{{r}_{std}}}{{P}_{r}} & 0\le\beta^{t,s}<{{r}_{c}}  \\
\displaystyle \frac{\beta^{t,s}}{{{r}_{std}}}{{P}_{r}} & {{r}_{c}}\le \beta^{t,s} < {{r}_{std}}  \\
{{P}_{r}} & \beta^{t,s} \ge {{r}_{std}}  \\
\end{array} \right.
\end{equation}
where $P_{PV}^{t,s}$ is the active PV power output; $\beta^{t,s}$ is the solar radiation measurement; ${{r}_{c}}$ is a certain radiation threshold level up to which a small increase in the radiation produces a significant increase in the PV output; ${{r}_{std}}$ is the solar radiation in the standard environment where further increase in radiation produces an insignificant change to the PV output; ${{P}_{r}}$ is the rated capacity of the solar installation. Considering that solar generation is typically injected at unity power factor \cite{WECC}, \color{black}PV active power is integrated in the EMS model whereas PV reactive power \color{black} is assumed to be zero in this study.

\subsection{ESS Modeling Constraints}
In this work, RB power is exploited as part of the ESS charging process. Following the approach in \cite{Sengor18, 8260299}, ESS modeling is formulated using the following constraints:
\begin{eqnarray}
\label{eq:ESS1}
 P_{RBE}^{t,s} + P_{B^+}^{t,s} &\leq& \bar{P}_{B+}u_{B}^{t,s} \quad  \forall t,s \\
 \label{eq:ESS2}
P_{B^-}^{t,s}&\leq& \bar{P}_{B-}(1-u_{B}^{t,s}) \quad  \forall t,s \\
 \label{eq:ESS0}
P_{B^+}^{t,s}, P_{B^-}^{t,s}&\geq& 0 \quad  \forall t,s \\
\label{eq:ESS3}
SoC_{B}^{t,s} &=& SoC_{B}^{t-1,s} - \epsilon_{B-}SoC_{B}^{t-1,s} \nonumber \\
\label{eq:ESS4}\hspace{30pt} & & + \hspace{3pt} \eta_{B+}(P_{RBE}^{t,s}+ P_{B^+}^{t,s})\Delta t \nonumber \\ & & - \hspace{3pt} \eta_{B-}P_{B^-}^{t,s}\Delta t \quad  \forall t,s\\
\label{eq:ESS5}
SoC_{B}^{t,s} &=& SoC_{B}^{0} \quad  \forall t=t_0\\
\label{eq:ESS6}
SoC_{B}^{t,s}  &\leq& SoC_{B}^{max} \quad  \forall t,s\\
\label{eq:ESS7}
SoC_{B}^{t,s}  &\geq& SoC_{B}^{min} \quad  \forall t,s
\end{eqnarray}
where $P_{B^+}^{t,s}$ is the ESS charging power; $P_{B^-}^{t,s}$ is the ESS discharging power; $SoC_{B}^{t,s}$ is the ESS energy level; $u_{B}^{t,s}$ is a binary variable to indicate the charging status of the ESS, i.e., its value is 1 during ESS charging and 0 during ESS discharging; $\epsilon_{B-}$ is the self-discharge coefficient; $\eta_{B+}$ is the ESS charging efficiency; $\eta_{B-}$ is the ESS discharging efficiency; $SoC_{B}^{0}$ is the initial ESS level \color{black}at $t_0$\color{black}; $SoC_{B}^{min}$ is the minimum ESS level; $SoC_{B}^{max}$ is the maximum ESS level. \color{black}Note that $SoC_B^0$ is typically expressed as a percentage of the ESS capacity and different percentage values corresponding to different ESS starting conditions may be selected\cite{Sengor18}.  \color{black}

\subsection{Regenerative Braking Energy Constraints}
To account for any limitations in the use of the available RB power due to the ESS charging capacity, the following constraint is added:
\begin{eqnarray}
\label{eq:rbelimits}
P_{RBE}^{t,s}&\leq& \bar{P}_{RBE}^{t,s}  \quad  \forall t, s
\end{eqnarray}
where $\bar{P}_{RBE}^{t,s}$ denotes the available RB power; $P_{RBE}^{t,s}$ denotes the RB power that is eventually utilized for the charging purposes of ESS.

\subsection{Power Balance Constraints}
An important condition of the EMS model is the satisfaction of the power balance constraint, i.e., the sum of power drawn from the main grid, renewable generation and any ESS discharging should meet the total demand of railway station loads including train demand, private and/or public EV charging and ESS charging. It is worth noting that a term should be integrated to account for any excess power that cannot be stored and may be sold back to the main grid. Hence, the power balance constraint is formulated as follows:
\begin{equation}
\label{eq:powerbalance}
P_G^{t,s}+P_{PV}^{t,s}+P_{B^-}^{t,s} = P_{D}^{t,s} + P_{B^+}^{t,s}+P_{EV}^{t,s}+P_{S}^{t,s} \quad  \forall t, s
\end{equation}
where $P_{D}^{t,s}$ is the train demand; $P_{EV}^{t,s}$ is the power demand for the charging of EVs at the train station.

\subsection{Power Exchange Constraints}
Although it is assumed that power can be bought from and sold back to the main grid, limitations on the amount of power that can be exchanged through the main grid should be considered. Moreover, the fact that power cannot be drawn and sold back to the main grid simultaneously should also be taken into consideration in the EMS model. Thus, the following constraints are included as part of the EMS model formulation:
\begin{eqnarray}
\label{eq:gridlimits}
P_{G}^{t,s}&\leq& \bar{P}_{G}u_{G}^{t,s} \quad  \forall t, s \\
\label{eq:gridlimits2}
P_{S}^{t,s}&\leq& \bar{P}_{S}(1-u_{G}^{t,s})  \quad  \forall t, s
\end{eqnarray}
where $\bar{P}_{G}$ is the maximum power that can be drawn from the main grid; $\bar{P}_{S}$ is the maximum power that can be sold back to the main grid; $u_{G}^{t,s}$ is a binary variable to indicate the direction of the power exchange between the train station and the main grid, i.e., its value is 1 if the train station draws power and 0 if it sells power back.

\section{An Optimal Energy Management Algorithm Integrating EV Charging}

\subsection{EV Charging Modeling}

We consider that EVs consist mainly of electric buses running from and to a specific railway station. Particularly, it is assumed that whenever a vehicle reaches the railway station, it charges until either its battery is fully charged or its next trip starts. On the other hand, once the vehicle leaves the station, discharging based on its specified route should be considered. The departure times for electric buses can be considered as known based on the public schedule. Hence, the EV battery energy level for each vehicle is as follows:
\begin{equation}
\label{eq:EVSoe}
SoC_{v}^{t} = SoC_{v}^{t-1} + \eta_{v+} P_{v^+}^{t}\Delta t - \eta_{v-}P_{v^-}^{t}\Delta t\\
\end{equation}
where $P_{v^+}^{t}, P_{v^-}^{t}$ are the charging and discharging power for vehicle $v$; $\eta_{v+}, \eta_{v-}$ are the charging and discharging efficiency, respectively. Charging and discharging can not happen at the same time.

In this work, EV charging demand at the station must be calculated and included as part of the power balance equation \eqref{eq:powerbalance}. Therefore, the set of plugged-in EVs to be charged at each time $t$ at the station is defined as:
\begin{equation}
\label{eq:pluggedinEVs}
\Omega_{EV}^t = \{ v: t_v^a \leq t < t_v^d, SoC_v^t<E_v^c \}
\end{equation}
where $t_v^a$ is the arrival time; $t_v^d$ is the departure time; $SoC_v^t$ is the EV battery energy level; $E_v^c$ is the required charging energy of vehicle $v$.

Thus, assuming charging with a constant nominal charging power $\bar{P}_v$, the overall EV demand at the railway station level can be estimated as:
\begin{equation}
\label{eq:estimatedpower}
P_{EV}^{t,s} = \sum_{v \in \Omega_{EV}^t} \bar{P}_{v}
\end{equation}

% Once the vehicle leaves the station for its specified route, discharging should be considered. Hence, the EV battery energy level for each vehicle is as follows:
% \begin{equation}
% \label{eq:EVSoe}
% SoC_{v}^{t} = SoC_{v}^{t-1} + \eta_{v+} P_{v^+}^{t}\Delta t - \eta_{v-}P_{v^-}^{t}\Delta t\\
% \end{equation}
% where $P_{v^+}^{t}, P_{v^-}^{t}$ are the charging and discharging power for vehicle $v$. Note that charging and discharging may not happen at the same time.

% However, considering the uncertainty of the departure times, it is possible that a customer may leave earlier than the fulfilment time. Therefore, a customer satisfaction threshold should be defined as follows:
% \begin{equation}
% \label{eq:satisfthreshold}
% \theta_v^k =  \begin{cases}
%   \eta_v \bar{P}_v \Delta t & \text{for } t_v^d < t_v^f \\
%   E_v^c-SoC_v^t & \text{for } t_v^d \geq t_v^f
%   \end{cases}
% \end{equation}

\subsection{The Proposed EMS Algorithm}
In this section, we describe a novel optimal EMS algorithm for a train station that integrates EV charging using the previously introduced models. The algorithm is summarized in \textbf{Algorithm \ref{alg:alg1}}. Particularly, \textbf{Steps 1-3} are for calculating the power consumed on EV charging, while \textbf{Steps 4-6} are for integrating the estimated EV charging requirements along with the available solar generation and solving the upper level EMS for the railway station.

\begin{algorithm}[!h]
\caption{The proposed Optimal EMS Algorithm}\label{alg:alg1}
\begin{algorithmic}
\STATE \hspace{-13.5pt} \underline{\textbf{EV Charging}}
\vspace{3pt}
\STATE \textbf{Step 1.} Given the arrival times, departure times and EV charging requirements, compute the set of plugged-in EVs $\Omega_{EV}^t$ to be charged at time $t$ using \eqref{eq:pluggedinEVs}. 
\STATE \textbf{Step 2.} Calculate the charging power needed at time $t$ using \eqref{eq:estimatedpower}.
\STATE \textbf{Step 3.} Move to the next time step $t=t+1$.
\vspace{5pt}
% \STATE \textbf{Step 3.}\hspace{0.2cm} \textbf{if} $\tilde{\lambda}<\hat{\lambda}$
% \STATE  \textbf{Step 4a.}\hspace{0.75cm} ${P_u^t}^\star=\min \{ \bar{P}_v, \frac{E_v^c-SoC_v^t}{ \eta_v \Delta t} \} $
% \STATE \hspace{1.35cm}\textbf{else} 
% \STATE \textbf{Step 4b.}\hspace{0.75cm} Solve the Receding Horizon Algorithm for \\
% \STATE \hspace{2.0cm}
% Optimal EV Charging using \eqref{mod:optevpolicy11}--\eqref{mod:optevpolicy7}.
% \vspace{5pt}
% \STATE \hspace{2cm} ${P_u^t}^\star={\text{min}}\quad && \lambda_p - \sum_{v \in \Omega_{EV}^t} \alpha_v^t  P_v^t$ 
% \vspace{3pt}
% \STATE \textbf{Step 4c.}\hspace{0.75cm}
% Update peak power due to EV charging
% \vspace{3pt}
% \STATE \hspace{2cm} $\hat{\lambda}=\sum_{v \in \Omega_{EV}^t} {P_v^t}^{\star} $
% \STATE \hspace{1.35cm}\textbf{end} 
\STATE \hspace{-13.5pt} \underline{\textbf{Optimal EMS Algorithm}}
\vspace{3pt}
\STATE \textbf{Step 4.} Given the available solar radiation measurements, calculate the predicted solar generated power using \eqref{eq:solar_p_val}.
\STATE \textbf{Step 5.} Apply the calculated EV charging requirements $P_{EV}^{t,s}$ and solar generation $P_{PV}^{t,s}$ in \eqref{eq:powerbalance}.
\STATE \textbf{Step 6.} Solve the EMS Optimization problem \eqref{eq:objfunction} \\
\STATE subject to constraints \eqref{eq:ESS1}--\eqref{eq:gridlimits2}. 
\end{algorithmic}
\label{alg1}
\end{algorithm}

\newcommand{\MATLAB}{\textsc{Matlab}\xspace}

\section{Numerical Results}
In this section, the case study to test the proposed optimal EMS Algorithm that integrates EV charging, RB, ESS and solar generation in electric railway stations is presented. 

\subsection{Simulation Setup}

\begin{figure}[!htb]
    \centering
    \includegraphics[width=2.1in,keepaspectratio=true,angle=0]{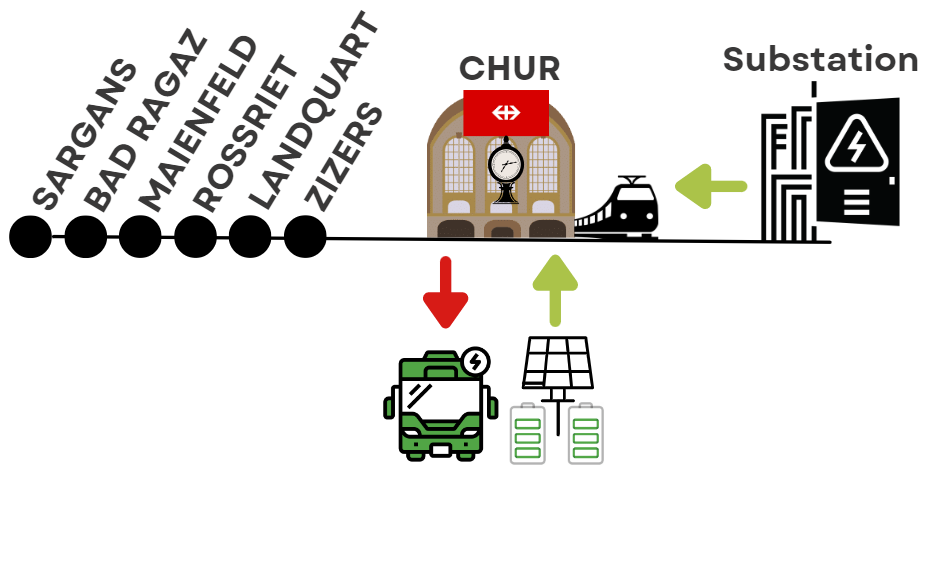}
    \caption{\vspace{-20pt}The route of Sargans-Chur railway line in Switzerland.}
    \vspace{-20pt}
    \label{route}
\end{figure}

The performance of the proposed method is tested on an actual railway line in Switzerland with a length of 24.9 km and 7 stations. An illustration of the selected route is shown in Fig. \ref{route}. To test the proposed EMS, Chur is selected as the main train station where the EV charging and PV generation are located. Actual daily train demand and RB data at 1-min resolution for the entire 2021 provided by the Swiss Federal Railways are utilized. ESS is assumed to be installed at the substation directly connected to Chur's train station.

Regarding EV demand, public electric buses at the bus stops closest to the main station are considered. EV arrival follows the public schedule on \cite{postauto}. The bus battery capacity $E_v^c$ is set as 280 kWh. The required charging power $\bar{P}_v$ is set as 300 kW. An average energy consumption of 0.87 kWh/min is assumed for the discharging of buses while on route.

% the parking lot at the train station is assumed to be open to EV arrivals daily from 6:00 to 22:00, whereas departure may happen either within or outside opening hours. For private EV cars, EV arrival follows an exponential distribution with an arrival rate of 4 vehicles per hour while the required charging energy follows a uniform distribution in the interval [0, 50] kWh. For public EV buses, EV arrival follows the public schedule as posted on \cite{postauto} whereas the required charging energy follows a uniform distribution in the interval [0, 300] kWh. 

Regarding solar generation, actual daily solar radiation data at 1-min resolution for 2021 are used and transformed into PV generated power using \eqref{eq:solar_p_val}, representing a solar penetration level of 20$\%$ of the peak train demand at the main train station. The associated parameters are selected as $r_c=150$ $\text{W/m}^2$ and $r_{std}=1000$ $\text{W/m}^\text{2}.$ 

Regarding ESS, it is assumed that it has a total capacity of $1000$ kWh based on \cite{esssbb}. The charging and discharging rates are set as $1000$ kW/min, the self-discharge coefficient is $\epsilon_{B-}=0$ and the charging and discharging efficiencies are $\eta_{B+}=\eta_{B-}=0.95$, respectively. The deep discharging limit $SoC_B^{min}$ is assumed as $10\%$, whereas the initial energy level $SoC_B^0$ is $50\%$ of the ESS capacity. 

Electricity prices consist of actual data of Switzerland's day-ahead market for 2021, as posted on \cite{entsoe}. It is assumed that the selling price $C_S^{t,s}$ and buying price $C_G^{t,s}$ are equal.

The study was implemented in the \textsc{Matlab}$^\copyright$ environment. The linear programming solver Gurobi \cite{gurobi} was used to obtain the optimal solution at \textbf{Step 6} of the proposed algorithm.

\vspace{-10pt}
\subsection{Validation of the Proposed Algorithm}
\label{casestudy1}
To validate the proposed EMS algorithm, we first focus on a single day. Particularly, the weekday with the highest amount of solar generation observed in the available data is selected. 

To evaluate the impact of different elements, four different cases are presented. \textit{Case 1} is considered as the base case where no ESS, RB or solar generation are included. In \textit{Case 2}, only ESS utilizing RB is added to the base case. In \textit{Case 3}, only solar generation is included additionally to the base case. Finally, \textit{Case 4} represents the full formulation, with ESS, RB and solar generation considered. 

\begin{figure}[!htb]
\centering
\subfloat[$\text{Train demand}$]{\includegraphics[width=1.695in]{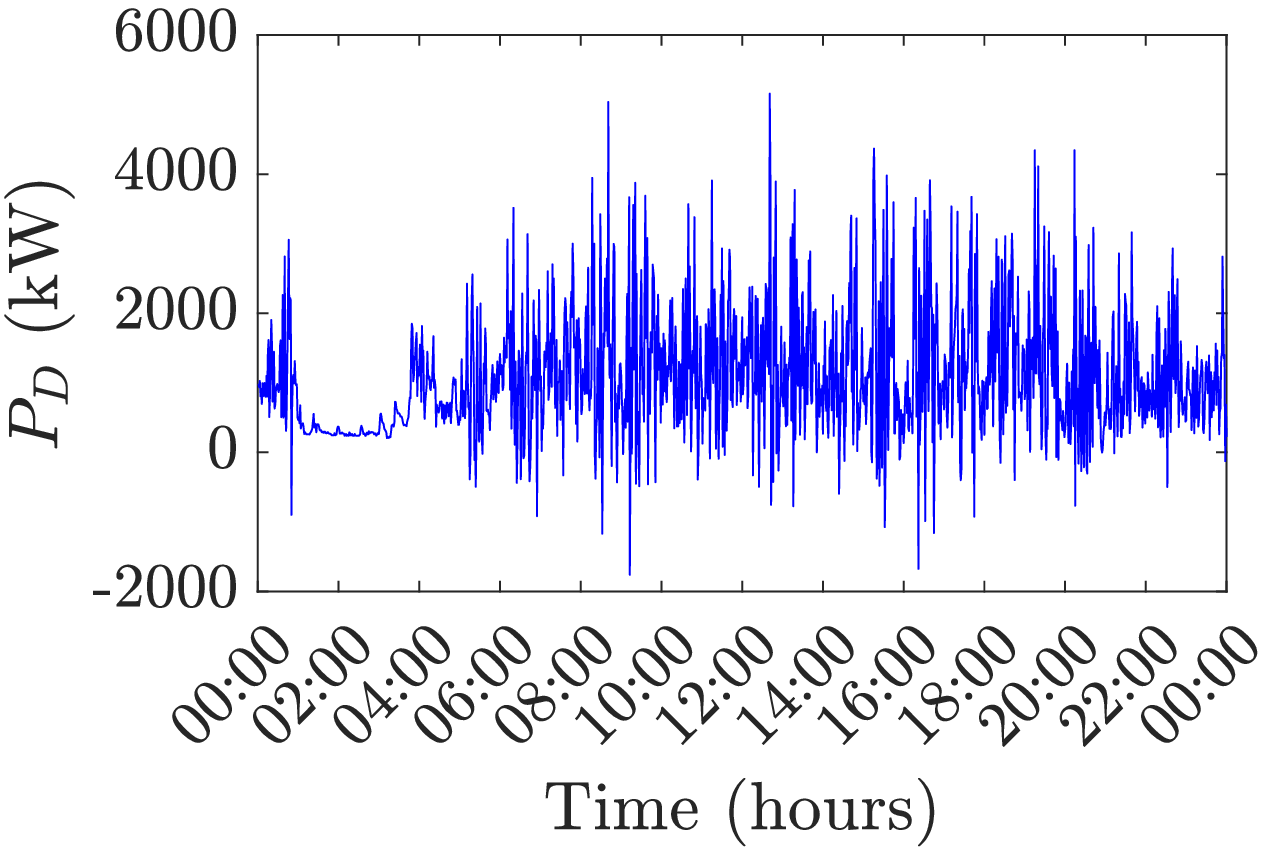}
\label{demand1}}
\hfil
\subfloat[EV demand]{\includegraphics[width=1.695in]{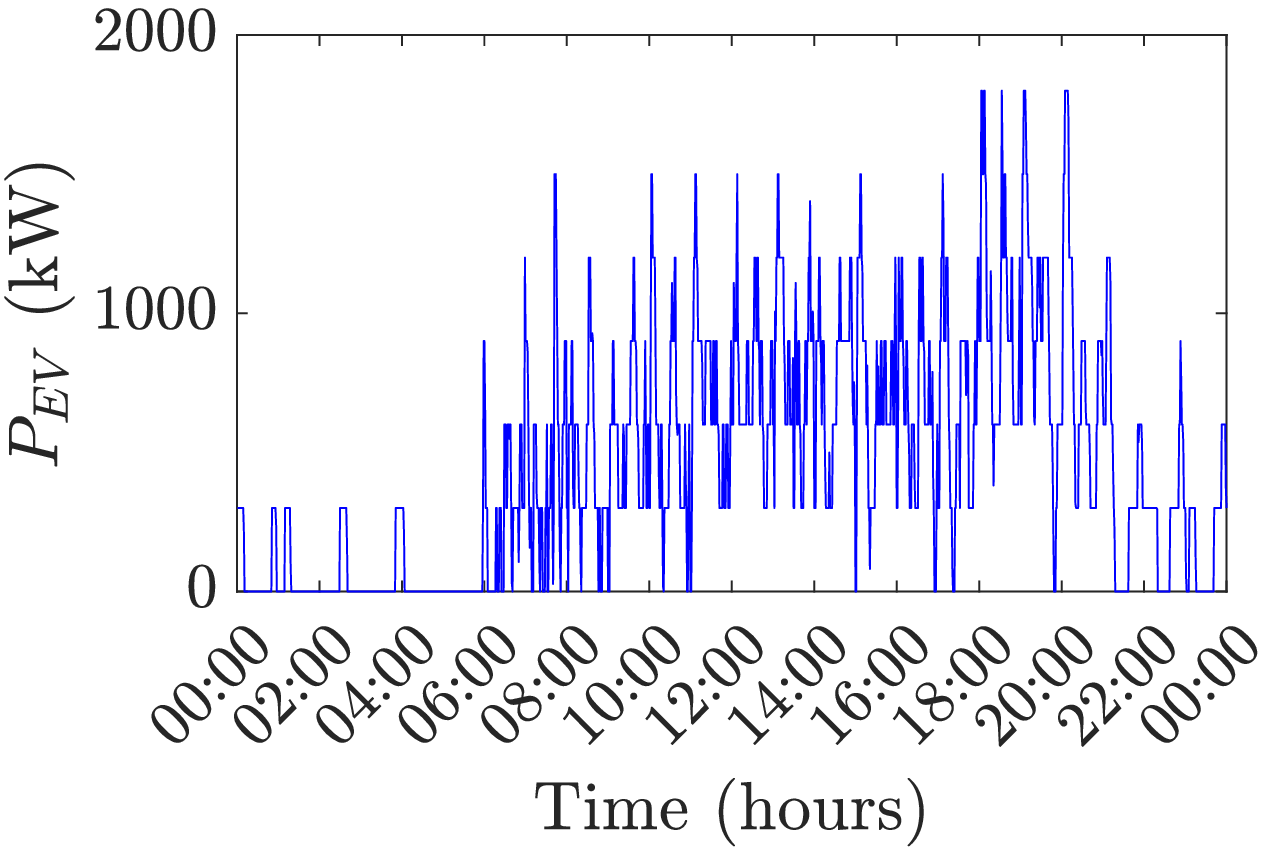}
\label{demand2}}
\caption{The considered daily power consumption profile at Chur station.} \label{demands} \vspace{-10.05pt}
\subfloat[RB Power]{\includegraphics[width=1.695in]{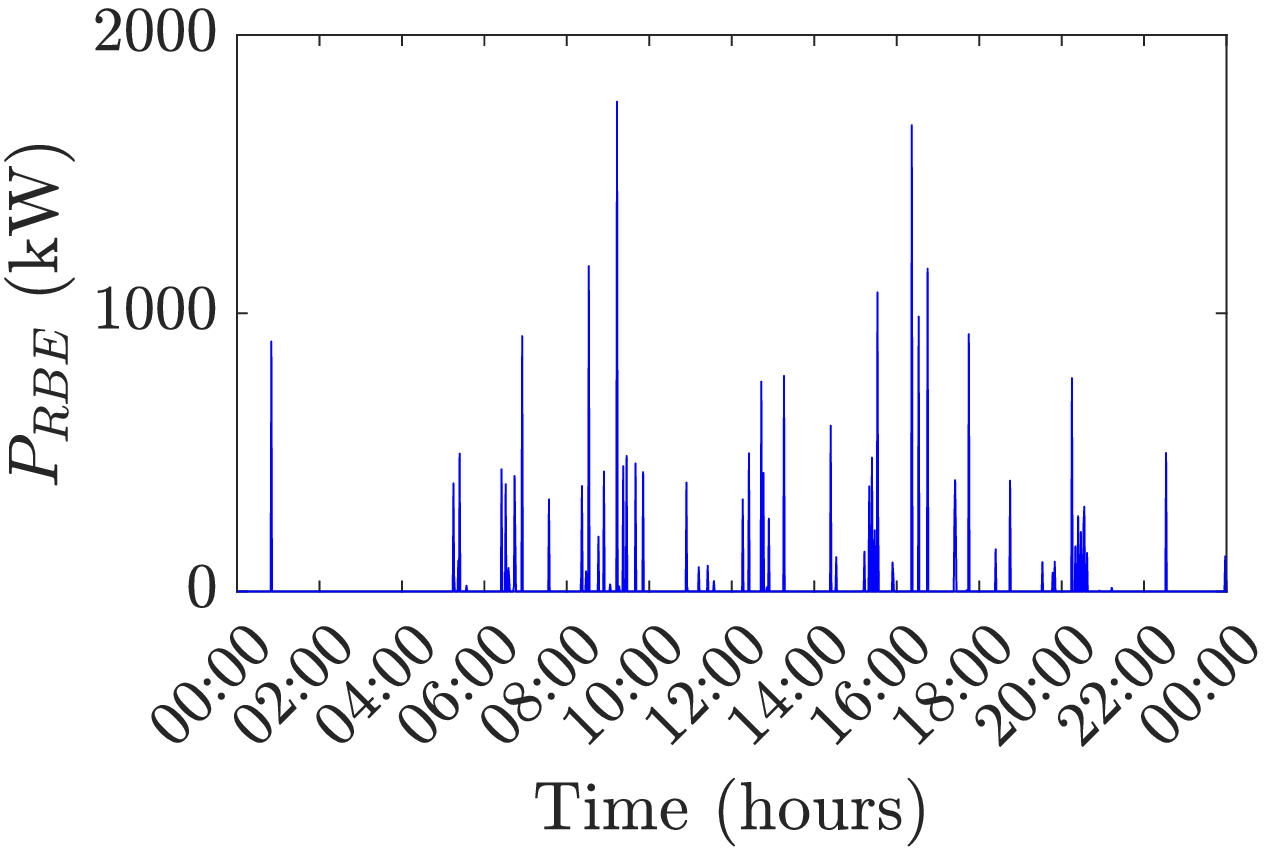}
\label{rbe1}}
\hfil
\subfloat[Day-ahead price]{\includegraphics[width=1.695in]{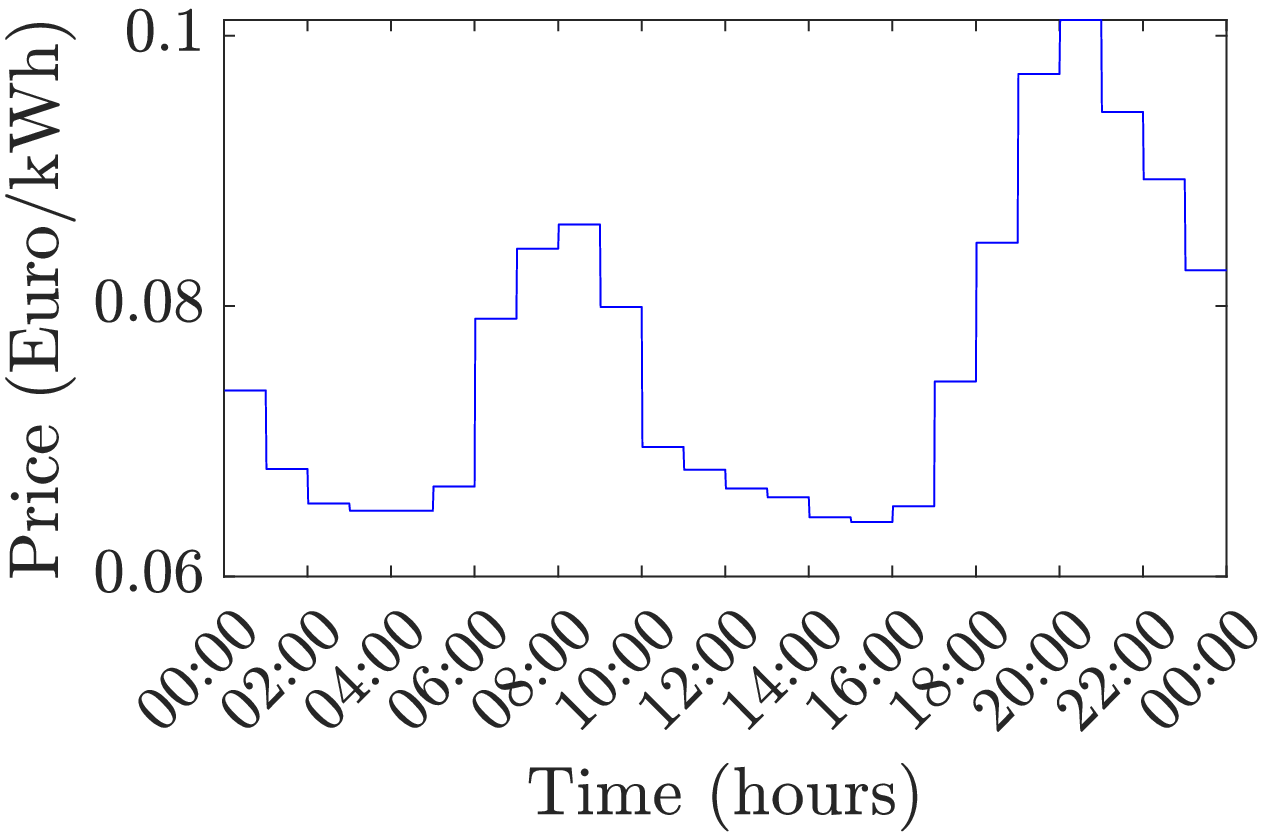}
\label{price2}}
\caption{The considered available RB power and day-ahead electricity price.}
\label{rbeprice}
\end{figure}

\begin{figure}[!ht]%[tb]
\centering
\includegraphics[width=2.4in ,keepaspectratio=true,angle=0]{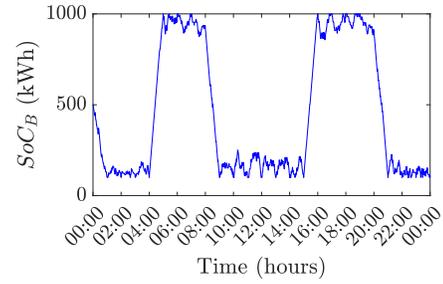}
\caption{The evolution of the ESS energy level for \textit{Case 4} of the selected day of interest.}
\label{fig:SOE1}
\vspace{-10pt}
\end{figure}

Figs. \ref{demands} and \ref{rbeprice} present the daily train demand, EV charging demand, RB availability and electricity price for the selected day of interest. Note that negative values in the train demand profile, as shown on Fig. \ref{demand1}, correspond to RB availability, which is further illustrated in Fig. \ref{rbe1}. In Fig. \ref{fig:SOE1}, the evolution of the state of energy for the ESS ($SoC_B^{t,s})$ for \textit{Case 4} of the selected scenario is presented. It can be observed that ESS stores energy between 04:00-05:00 despite the lack of RB availability, which is due to the low electricity price. Moreover, ESS is discharged to cover the combined peak of EV charging and train demand between 08:00-09:00 when the price increases. No major ESS use is observed between 09:00-15:00, since solar generation is mostly available in this timeframe and can be directly used to cover part of the demand. Later, thanks to RB availability and low electricity prices, ESS is charged between 15:00-16:00. Finally, ESS full discharging takes place between 20:00-21:00, as a reaction to the highest observed electricity price throughout the day. 

Table \ref{table1} presents the daily operating costs for each of the considered cases. It can be observed that \textit{Case 4} that coordinates train demand, EV charging, RB, ESS and solar generation is the most profitable, leading to cost savings up to 25.93$\%$. The results are aligned with the theoretical expectation that the deployment of ESS and solar generation for serving the overall station consumption greatly contributes to cost savings. The level of savings is of course dependent on the specific input data, i.e., demand, prices, and solar irradiation. 
\vspace{-23pt}

% \begin{figure}[!htb]
% \centering
% \subfloat[$\text{Train demand}$]{\includegraphics[width=1.695in]{train_day165_.eps}
% \label{demand1}}
% \hfil
% \subfloat[EV demand]{\includegraphics[width=1.695in]{EV_1min_ticks_larger_.eps}
% \label{demand2}}
% \caption{The considered daily power consumption profile at Chur station.}
% \label{demands}
% \end{figure}

% \vspace{-20pt}
% \vspace{-28pt}
% \begin{figure}[!htb]
% \centering
% \subfloat[RB Power]{\includegraphics[width=1.695in]{rbe_day165_.eps}
% \label{rbe1}}
% \hfil
% \subfloat[Day-ahead price]{\includegraphics[width=1.695in]{price_day165.eps}
% \label{price2}}
% \caption{(a) The available RB power.
% (b) The day-ahead electricity price for the selected day of interest. }
% \label{rbeprice}
% \end{figure}

% \begin{figure}[!t]%[tb]
% \centering
% \includegraphics[width=2.5in ,keepaspectratio=true,angle=0]{pluggedin_EVs.eps}
% \caption{Number of plugged-in EVs.}
% \label{fig:pluggedEV}
% \end{figure}

% \begin{figure}[!ht]%[tb]
% \centering
% \includegraphics[width=2.5in ,keepaspectratio=true,angle=0]{train_1min_ticks.eps}
% \caption{Train demand.}
% \label{fig:traindemand}
% \end{figure}

% \begin{figure}[!ht]%[tb]
% \centering
% \includegraphics[width=2.5in ,keepaspectratio=true,angle=0]{EV_1min_ticks.eps}
% \caption{EV charging demand.}
% \label{fig:EVdemand}
% \end{figure}

% \begin{figure}[!ht]%[tb]
% \centering
% \includegraphics[width=2.5in ,keepaspectratio=true,angle=0]{price_1min_ticks.eps}
% \caption{Day-ahead price.}
% \label{fig:price}
% \end{figure}

% \begin{figure}[!ht]%[tb]
% \centering
% \includegraphics[width=2.5in ,keepaspectratio=true,angle=0]{RBE_1min_ticks.eps}
% \caption{Regenerative Braking Energy.}
% \label{fig:rbe}
% \end{figure}

\vspace{-10pt}

\begin{table}[!htb]
\centering
  \caption{Results - 1 Scenario}\label{table1}
   \setlength{\tabcolsep}{2pt}
  \begin{tabular}{|c|c|c|c|c|}
\hhline{|-|-|-|-|-|}
\textbf{Case}& \textbf{ESS} & \textbf{PV} & \textbf{Total Costs (€)} & \textbf{Cost Savings} (\%)
\\
\hline
1&-&-&2861.96&-
\\
\hline
2&\checkmark&-&2742.96&4.16\\
\hline
3&-&\checkmark&2238.97&21.77 \\
\hline
4&\checkmark&\checkmark&2119.96&25.93\\
\hhline{|-|-|-|-|-|}
  \end{tabular}
  
\end{table}

\subsection{A Scenario-Based Approach}

In this section, multiple scenarios are utilized to validate the proposed EMS algorithm and compute the expected operating cost over the year while taking into account variations in input data. Specifically, 200 scenarios of equal probability $\pi_s$ corresponding to different days, solar radiation, and train demand profiles are tested. \color{black}Note that the uncertainty path realization is based on historical data of year 2021 provided by the Swiss Federal Railways or available online \cite{entsoe}.\color{black} 

Similarly to Section \ref{casestudy1},  Table \ref{table2} presents the results for \textit{Case 1, Case 2, Case 3} and \textit{Case 4}. The proposed \textit{Case 4}, integrating ESS, RB and solar generation for the combined railway and EV charging demand, is still the most profitable case across multiple scenarios with achieved cost savings of up to 17.81$\%$, indicating the efficiency of the proposed EMS model. Additionally, it can be observed that when multiple scenarios are considered, the expected value of the daily operating cost increases comparing to the daily costs in Table \ref{table1}, where the focus was only on 1 scenario of the day with the highest solar generation observed.  Intuitively, the larger
the amount of solar generation is, the smaller the daily operating cost is. Indeed, such results from this
systematic scenario-based approach are consistent with our previous conjecture.

\begin{table}[!b]
\centering
\vspace{-10pt}
  \caption{Results - 200 Scenarios}\label{table2}
   \setlength{\tabcolsep}{2pt}
  \begin{tabular}{|c|c|c|c|c|}
\hhline{|-|-|-|-|-|}
\textbf{Case}& \textbf{ESS} & \textbf{PV} & \textbf{Total Costs (€)} & \textbf{Cost Savings} (\%)
\\
\hline
1&-&-&3226.71&-
\\
\hline
2&\checkmark&-&3036.08&5.91\\
\hline
3&-&\checkmark&2842.81&11.90 \\
\hline
4&\checkmark&\checkmark&2652.16&17.81\\
\hhline{|-|-|-|-|-|}
  \end{tabular}
  
\end{table}

\begin{figure}[!ht]%[tb]
\centering
\includegraphics[width=2.4in ,keepaspectratio=true,angle=0]{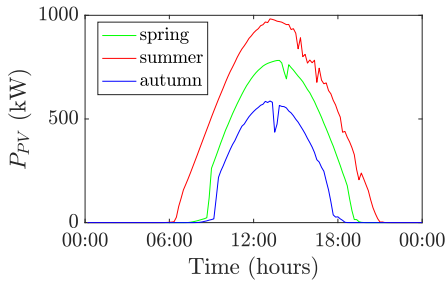}
\caption{The solar generated power over different scenarios.}
\label{fig:comparison1}
\end{figure}

\begin{figure}[!ht]%[tb]
\centering
\includegraphics[width=2.5in ,keepaspectratio=true,angle=0]{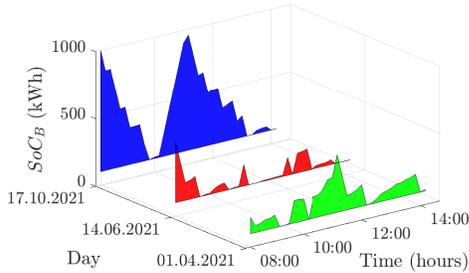}
\caption{The evolution of the ESS energy level over different scenarios.}
\label{fig:comparison2}
\end{figure}

To further illustrate the results of the scenario-based approach, Fig. \ref{fig:comparison1} pictures a comparison of the PV generation for three days, including a spring, summer, and autumn day. The summer day corresponds to the selected scenario presented in Section \ref{casestudy1}. It can be seen that solar availability may greatly vary across scenarios, as a $\sim$50$\%$ decrease in terms of peak PV power as well as different duration are observed when comparing the PV generation on summer and autumn days. Such variability in input data may influence the overall EMS operation. Indeed, Fig. \ref{fig:comparison2} shows the ESS behavior with respect to the different scenarios. Looking at the autumn and summer days, despite the fact that the combined railway and EV charging demand profiles as well as the day-ahead prices follow similar patterns during the peak hours 08:00-10:00, ESS discharges in the autumn case to serve the combined load as solar is not available during this time period. Similar behavior is observed during 11:00-13:00, as existing PV generated power may not be sufficient. Hence, we can conclude that ESS activation greatly varies to meet the peaks at the train and EV charging load due to input data variations, highlighting the need for scenario-based approaches.

\vspace{-11pt}

% \begin{figure}[!ht]%[tb]
% \centering
% \includegraphics[width=2.6in ,keepaspectratio=true,angle=0]{soe_comparison_samecolor.eps}
% \caption{The evolution of the ESS energy level over different scenarios.}
% \label{fig:comparison2}
% \end{figure}

\vspace{-15pt}

% \begin{figure}[!ht]%[tb]
% \centering
% \includegraphics[width=2.6in ,keepaspectratio=true,angle=0]{soe_comparison_face075.eps}
% \caption{The evolution of the ESS energy level over different scenarios.}
% \label{fig:comparison2}
% \end{figure}
\vspace{-10pt}
\section{Conclusion}
Leveraging the electric railway infrastructure, this
paper proposes a novel optimal EMS algorithm considering RBE, ESS and renewable generation for the combined train and EV charging demand at electric railway stations.
Specifically, the proposed approach estimates the charging requirements for the electric buses serving the closest stops to the station and effectively coordinates the available RBE, ESS and conventional and PV generation to minimize the daily operating cost. 
The proposed method represents the first attempt to meet the combined train and EV charging demand of electric railway stations and unlike previous works, considers the potential of selling any excess power back to the grid as well as variations in the input data, such as renewable generation. Numerical results on an actual railway line in Switzerland show that the proposed EMS method can provide effective coordination of the EV charging while ensuring optimal operation and cost savings across a large number of scenarios. Indeed, cost savings up to 25.93$\%$ with respect to the base case were observed.
Future work may focus on extending the results to account for EV charging flexibility and \color{black} include other uncertainties in the set of scenarios\color{black}, such as the initial energy level for the ESS.

\section*{Acknowledgments}
The authors would like to thank Robert Strietzel and the Swiss Federal Railways for providing the railway consumption and solar radiation data to test the proposed algorithm.

% \appendices{}
% \section{}
% % {\appendix[A]
% % }


% \section{References Section}
% You can use a bibliography generated by BibTeX as a .bbl file.
%  BibTeX documentation can be easily obtained at:
%  http://mirror.ctan.org/biblio/bibtex/contrib/doc/
%  The IEEEtran BibTeX style support page is:
%  http://www.michaelshell.org/tex/ieeetran/bibtex/
 
%  % argument is your BibTeX string definitions and bibliography database(s)
% %\bibliography{IEEEabrv,../bib/paper}
% You can manually copy in the resultant .bbl file and set second argument of $\backslash${\tt{begin}} to the number of references
%  (used to reserve space for the reference number labels box).
%========== Bibliography

\bibliographystyle{IEEEtran}
\bibliography{pesgm_review_black}

\vfill

\end{document}